\documentclass[preprint,11pt,authoryear]{elsarticle}
\usepackage{amssymb}
\usepackage{lipsum}

\journal{Icarus}
\usepackage{xcolor}
\usepackage{makecell} 
\usepackage[colorlinks=true, linkcolor=blue, citecolor=blue, urlcolor=blue]{hyperref}
\usepackage{academicons}
\usepackage{url}
\usepackage[letterpaper, left=3.5cm, right=3.5cm, top=3.5cm, bottom=3.5cm]{geometry}

\begin{document}
\begin{frontmatter}
\title{Spectral Mixture Modeling with Laboratory Near-Infrared Data I: Insights into Compositional Analysis of Europa}

\author[first]{A. Emran}
\affiliation[first]{organization={NASA Jet Propulsion Laboratory},
            addressline={California Institute of Technology}, 
            city={Pasadena},
            postcode={91109}, 
            state={CA},
            country={USA}}
\ead{al.emran@jpl.nasa.gov}
\begin{abstract}
Europa's surface composition and physical characteristics are commonly constrained using spectral deconvolution through linear mixture (LM) modeling and radiative transfer–based (RT) intimate mixture modeling. Here, I compared the results of these two spectral modeling— LM versus RT— against laboratory spectra of water (H\textsubscript{2}O) ice and sulfuric acid octahydrate (SAO; H\textsubscript{2}SO\textsubscript{4}·8H\textsubscript{2}O) mixtures measured at near-infrared wavelengths ($\sim$1.2–2.5 $\mu$m) with grain sizes of 90–106 $\mu$m \citep{hayes2025insights}. The modeled abundances indicate that the RT more closely reproduces the laboratory abundances, with deviations within ±5\% for both H\textsubscript{2}O ice and H\textsubscript{2}SO\textsubscript{4}$\cdot$8H\textsubscript{2}O with $\sim$100 $\mu$m grains. In contrast, the LM shows slightly larger discrepancies, typically ranging from ±5-15\% from the true abundances. Interestingly, both LM and RT tend to consistently overestimate the abundance of H\textsubscript{2}SO\textsubscript{4}·8H\textsubscript{2}O and underestimate H\textsubscript{2}O ice across all mixtures. Nonetheless, when H\textsubscript{2}SO\textsubscript{4}$\cdot$8H\textsubscript{2}O either dominates \citep[\(>80\%\) as observed on Europa’s trailing hemisphere;][] {carlson2005distribution} or is present only in trace amounts \citep[$\sim$10\% on areas in Europa’s leading hemisphere;][]{dalton2013exogenic, ligier2016vlt}, both the LM and RT render acceptable results within ±10\% uncertainty. Thus, spectral modeling using the RT is preferred for constraining the surface composition across Europa, although the LM remains viable in specific compositional regimes.
\end{abstract}

\end{frontmatter}

\section{Introduction}
\label{introduction}
Estimations of planetary surface composition—on both terrestrial and icy bodies— across the Solar System are done using spectral deconvolution via linear mixture (LM) modeling and radiative transfer (RT) theory, known as intimate mixture modeling \citep[e.g.,][]{li2011radiative, dalton2012europa, li2015estimating, goudge2015integrating, shirley2016europa, liu2016end, emran2021thermophysical, emran&stack2025, emran2025nh}. In linear mixture modeling, also commonly known as the areal mixture approach, the reflectance of a mixture is represented as a linear combination of the spectra of its constituent materials, assuming that each material exists in spatially distinct patches \citep[e.g.,][]{hapke2012theory, dalton2012europa, emran2023surface, stack2015modeling}. In contrast, radiative transfer-based intimate mixture modeling assumes that the component materials are physically mixed in a “salt-and-pepper” fashion, resulting in a nonlinear combination of the reflectance spectra of the constituents \citep[e.g.,][]{mustard1987quantitative,poulet2004nonlinear, li2011radiative, hapke2012theory, li2015estimating}. In planetary science, Hapke’s radiative transfer model \cite{hapke1981bidirectional, hapke2001space, hapke2012theory} has been extensively applied for spectral modeling to constrain the compositions of icy bodies in the outer Solar System \citep[refer to][for exhaustive references]{emran&chevrier2022, emran&chevrier2023, khuller2025quantitative}.

Although both the LM and RT spectral modeling have their inherent advantages and limitations \citep[e.g.,][]{keshava2002spectral, hapke2012theory}, they have been successfully applied to constrain the surface composition of Europa \citep[e.g.,][]{shirley2016europa}. Using data from both ground- and space-based observations, Europa’s surface composition has been derived from visible and near-infrared (VIS-NIR) wavelengths using the LM modeling \citep[e.g.,][]{dalton2007linear, dalton2012europa, shirley2010europa, brown2013salts, ligier2016vlt, davis2024pwyll, emran2025nh} and the RT modeling based on \cite{hapke1981bidirectional} theory \citep[e.g.,][]{carlson2005distribution, mishra2021comprehensive, mermy2023selection}. While a variety of ice and non-ice components have been hypothesized on Europa’s surface, water (H\textsubscript{2}O) ice and sulfuric acid octahydrate (SAO; H\textsubscript{2}SO\textsubscript{4}·8H\textsubscript{2}O) have consistently been identified as the major constituents using both the LM and RT modeling approaches, as well as by comparisons with laboratory spectra \citep[e.g.,][]{Moroz1966, Pilcher1972, carlson1999sulfuric, carlson2005distribution, dalton2007linear, dalton2012europa, shirley2010europa, shirley2016europa, ligier2016vlt, mermy2023selection, emran2025nh}.

Nonetheless, evaluating the accuracy of the LM and RT modeling in constraining surface composition remains a critical area of research for accurately understanding Europa’s surface and interpreting its geology. \cite{shirley2016europa} compared abundance estimates derived from the LM and RT modeling using data from the Galileo Near-Infrared Mapping Spectrometer \citep[NIMS;][]{carlson1992near}. Their analysis of spectra from various regions on Europa revealed that both approaches rendered consistent results, with linear correlation coefficients of $\ge$ 0.98 for H\textsubscript{2}O ice and H\textsubscript{2}SO\textsubscript{4}$\cdot$8H\textsubscript{2}O abundances by both models \citep{shirley2016europa}. However, until this current study, validation of these mixture modeling using laboratory data has remained lacking. Recently, \cite{hayes2025insights} measured reflectance spectra of H\textsubscript{2}O ice and H\textsubscript{2}SO\textsubscript{4}·8H\textsubscript{2}O mixtures at varying ratios relevant to Europa— offering a unique opportunity to assess spectral modeling accuracy. In this study, as part of a series of investigations \citep{emran2025paper2}, I assess the accuracy of the LM and RT modeling against the laboratory spectra of H\textsubscript{2}O ice and H\textsubscript{2}SO\textsubscript{4}$\cdot$8H\textsubscript{2}O mixtures at near-infrared (NIR) wavelengths— providing a framework for quantitatively constraining Europa’s surface composition.

\section{Observation and Methods}
\subsection{Laboratory spectra}
\begin{figure*}
	\centering 
	\includegraphics[width=1.\textwidth, angle=0]{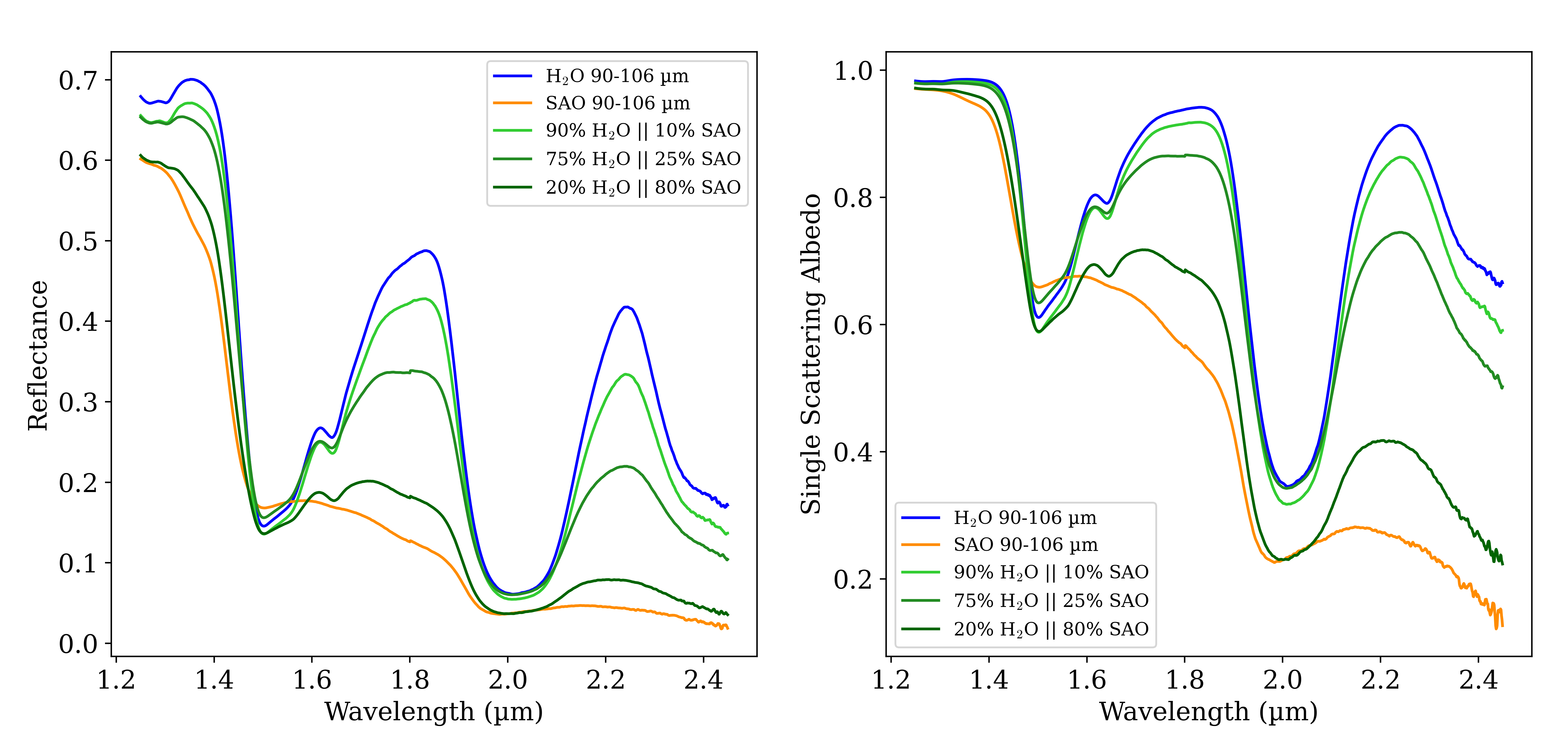}	
	\caption{\textit{Left panel:} Laboratory reflectance spectra of H\textsubscript{2}O ice, H\textsubscript{2}SO\textsubscript{4}$\cdot$8H\textsubscript{2}O (sulfuric acid octahydrate; SAO), and their mixtures at varying proportions (\%wt) and grain sizes of 90–106 $\mu$m \citep{hayes2025insights}. The blue spectrum represents the H\textsubscript{2}O ice endmember, the orange spectrum represents the H\textsubscript{2}SO\textsubscript{4}$\cdot$8H\textsubscript{2}O endmember, and the spectra shown in shades of green correspond to mixtures of H\textsubscript{2}O ice and H\textsubscript{2}SO\textsubscript{4}$\cdot$8H\textsubscript{2}O at different ratios. All reflectance spectra were collected from \cite{hayes2025insights}. \textit{Right panel:} Single scattering albedo spectra corresponding to the reflectance spectra, derived using radiative transfer theory based on the \cite{hapke1981bidirectional} model (refer to \textit{Section 2.2} for conversion details).} 
	\label{fig1}%
\end{figure*}

Reflectance spectra of H\textsubscript{2}O ice and H\textsubscript{2}SO\textsubscript{4}·8H\textsubscript{2}O mixtures were measured at varying ratios and in the grain sizes of 90–106 $\mu$m \citep{hayes2025insights}. The measurements were conducted at $\sim$77 K under standard viewing geometry, with an incidence angle (\textit{i}) of 30°, an emission angle (\textit{e}) of 0°, and a phase angle (\textit{g}) of 30° at the spectral wavelength of $\sim$1.25-2.45 $\mu$m. Although Europa’s surface temperature is $\sim$120 K \citep[e.g.,][]{spencer1999temperatures, carlson2009europa, ashkenazy2019surface}, reflectance spectra of H\textsubscript{2}O ice and H\textsubscript{2}SO\textsubscript{4}$\cdot$8H\textsubscript{2}O measured at 77 K can be used to represent Europa’s surface relevant spectra, as spectral variations (absorption depths and position) over this temperature range are minor \citep[e.g.,][]{carlson2005distribution, mastrapa2008optical, mastrapa2009optical,stephan2021vis, hayes2025insights}. For details on the experimental setup and procedures, I refer readers to \cite{hayes2025insights}. The spectra of pure H\textsubscript{2}O ice and pure H\textsubscript{2}SO\textsubscript{4}$\cdot$8H\textsubscript{2}O obtained by \cite{hayes2025insights} are used as endmembers for both the LM and RT spectral modeling. The mixture dataset includes three compositions: 90\% H\textsubscript{2}O + 10\% SAO, 75\% H\textsubscript{2}O + 25\% SAO, and 20\% H\textsubscript{2}O + 80\% SAO mixtures (by mass or \%wt) at grain sizes of $\sim$100 (90–106) $\mu$m (Fig. \ref{fig1}). The H\textsubscript{2}SO\textsubscript{4}$\cdot$8H\textsubscript{2}O grains used in this study are spherical, whereas the H\textsubscript{2}O ice grains are angular to sub-angular (T. W. Hayes, 2025; personal communication).

\subsection{LM and RT modeling}

In linear mixture modeling, the mixture spectrum is represented as the weighted sum of the reflectances of each endmember, with the weights corresponding to their abundances \citep[e.g.,][]{emran2023surface}:

\begin{equation}
    r_m = \sum_{i=1}^{k} f_i \cdot r_i ~ ; ~  0 \leq f_i \leq 1 ~; ~\sum_{i=1}^{k} f_i = 1
    \label{eq1}
\end{equation}

where \textit{r\textsubscript{m}} is the reflectance of the mixture spectrum, \textit{r\textsubscript{i}} is the reflectance of the \textit{i\textsuperscript{th}} constituent endmember, and \textit{f\textsubscript{i} }is the fractional abundance (\%wt) of the \textit{i\textsuperscript{th}} endmember \citep[e.g.,][]{dalton2012europa, emran2021thermophysical, emran&stack2025}. Using Eq. (\ref{eq1}), I estimated the modeled abundances of H\textsubscript{2}O ice and H\textsubscript{2}SO\textsubscript{4}$\cdot$8H\textsubscript{2}O for reflectance spectra of mixtures. 

In contrast, intimate mixture modeling based on radiative transfer accounts for the non-linear scattering properties of reflectance, but assumes linear behavior when reflectance is converted to single scattering albedo \citep{hapke1981bidirectional, hapke2001space, hapke2012theory, mustard1989photometric,mustard2019theory}. Therefore, I converted both the endmember and mixture reflectance spectra to single scattering albedo using the Hapke theory \citep{hapke1981bidirectional} with the same viewing geometry and grain size as the measured reflectance spectra (Fig. \ref{fig1}). The reflectance (also known as radiance coefficient) in the Hapke model \citep{hapke1981bidirectional} can be expressed as:
\begin{equation}
r(i, e, g) = \frac{w}{4} \cdot \frac{1}{\mu_e + \mu_i} 
\left\{ \left[1 + B(g)\right] P(g) + H(\mu_e) H(\mu_i) - 1 \right\}
\label{eq:2}
\end{equation}

where \textit{r} is the reflectance, \textit{w} is the average single-scattering albedo\textit{, $\mu$\textsubscript{i}} and \textit{ $\mu$\textsubscript{e}} are the cosines of incidence angle (\textit{i}) and emission angle (\textit{e}), respectively, \textit{B}(\textit{g}) is the backscattering function, \textit{P}(\textit{g}) is the single-particle phase function, and \textit{H($\mu$\textsubscript{x})} is the multiple scattering \textit{H} functions of \cite{chandrasekhar2013radiative}. Refer to relevant literature for details of these functions \citep[e.g.,][]{hapke1981bidirectional, hapke2001space, hapke2012theory}. However, for the laboratory spectra considered in this study, I assume no backscattering (\textit{B}(\textit{g}) = 0) and an isotropic phase function (\textit{P(g) = 1}), following \cite{lapotre2017probabilistic}. Under these assumptions, the Hapke reflectance equation simplifies to \citep[e.g.,][]{mustard1987quantitative,mustard1989photometric}:

\begin{equation}
r(i, e, g) \approx \frac{w}{4} \cdot \frac{1}{\mu_e + \mu_i} \cdot H(\mu_e) H(\mu_i)
\label{eq:3}
\end{equation}

The validity of Eq. \ref{eq:3} for abundance estimation has been tested against laboratory reflectance spectra of intimate mineral mixtures, under the assumption of isotropic scattering behavior of particulate media at phase angles of 15° – 40° \citep{mustard1987quantitative, mustard1989photometric}. At these intermediate phase angles, the backscattering is assumed negligible (\textit{B(g)} \(<\)\(<\) 1) and becomes significant only at very small phase angles (\(<10\)°; \citealt{mustard1987quantitative}). Using this simplified method (Eq. \ref{eq:3}), abundance estimates are accurate to within a few percent (\%) when particle grain sizes are known \citep{mustard1987quantitative, mustard1989photometric}.

The Chandrasekhar \textit{H} functions describing multiple scattering can be derived as \citep{hapke2002bidirectional}:

\begin{equation}
H(x) \approx \left[ 1 - wx \left( r_o + \frac{1 - 2 r_o x}{2} \ln\left( \frac{1 + x}{x} \right) \right)^{-1} \right]
\label{eq:4}
\end{equation}

\begin{equation}
r_o = \frac{1 - \gamma}{1 + \gamma}
\label{eq:5}
\end{equation}

\begin{equation}
\gamma = \sqrt{1 - w}
\label{eq:6}
\end{equation}

where \textit{r\textsubscript{o}} is the bihemispherical reflectance for isotropic scatterers, $\gamma$  is the albedo factor, and \textit{x} represents either the cosine of the incidence angle ($\mu$\textsubscript{i} = cos(\textit{i})) or the emission angle ($\mu$\textsubscript{e} = cos(\textit{e})). 

The photon penetration depth (\textit{$\delta$}) into a particulate surface or media is inversely proportional to the imaginary part (\textit{k}) of the material's optical constant at a given wavelength \citep[e.g.,][]{born1999principles, hapke2012theory}. For both H\textsubscript{2}O ice and H\textsubscript{2}SO\textsubscript{4}$\cdot$8H\textsubscript{2}O, the low \textit{k} values at NIR  \citep[e.g.,][]{mastrapa2008optical, mastrapa2009optical, carlson2005distribution} result in \textit{$\delta$} on the order of hundreds of micrometers. Because the grain diameter (\textit{d} $\sim$100$\mu$m) is much smaller than the penetration depth (\textit{d} \(<\) \textit{$\delta$}), photons interact with multiple grains, making multiple scattering the dominant effect \citep[e.g,][]{hapke1981bidirectional, hapke2012theory}. This process can induce significant spectral changes in intimate mixture even with trace amounts of material \citep[e.g.,][]{hayes2025insights}. Thus, multiple scattering terms (\textit{H}-functions) were included in RT modeling (Eq. \ref{eq:3}) to ensure accurate abundance estimates. 

Upon conversion to single scattering albedo from the laboratory reflectance spectra using Eqs. (\ref{eq:3})–(\ref{eq:6}), the RT spectral modeling can be expressed as follows \citep[e.g.,][]{stack2015modeling, goudge2015integrating, lapotre2017probabilistic}:

\begin{equation}
w_m = \sum_{i=1}^{k} f_i \cdot w_i
\label{eq:7}
\end{equation}

where \textit{w\textsubscript{m}} is the single scattering albedo of the mixture spectrum, \textit{w\textsubscript{i}} is the single scattering albedo of the \textit{i\textsuperscript{th}} constituent endmember, and \textit{f\textsubscript{i} }is the relative fractional (geometric) cross section of the \textit{i\textsuperscript{th}} endmember. The relative geometric cross-section (also known as F-parameter) is a function of the mass fraction, density, and particle diameter of components in a mixture \citep[e.g.,][]{mustard1987quantitative, mustard1989photometric}. 

While grain size has a major influence on the spectral characteristics of materials, grain shape can also affect the resulting spectra \citep[e.g.,][]{Shkuratov2005, hapke2012theory}. However, since both endmembers— H\textsubscript{2}O ice and H\textsubscript{2}SO\textsubscript{4}$\cdot$8H\textsubscript{2}O —were modeled with the same grain size ($\sim$100$\mu$m), shape effects were ignored. Under simplified assumptions, such as similar grain sizes and densities of particles, \textit{f\textsubscript{i}} can be considered equivalent to the fractional contribution or abundance (\%wt) of the corresponding endmember \citep[e.g.,][]{mustard1987quantitative}. However, the densities of H\textsubscript{2}O ice and SAO differ slightly under outer Solar System–relevant temperatures— H\textsubscript{2}O ice has a density of $\sim$1 g/cm\textsuperscript{3}  \citep[e.g.,][]{blackman1957cubic, yarnall2022crystalline}, whereas the slightly denser H\textsubscript{2}SO\textsubscript{4}$\cdot$8H\textsubscript{2}O has a density of $\sim$1.3 g/cm\textsuperscript{3} \citep{maynard2012structure, maynard2013new}. Accordingly, the relative fractional cross-section for H\textsubscript{2}O ice and H\textsubscript{2}SO\textsubscript{4}$\cdot$8H\textsubscript{2}O was scaled by their respective density factors in Eq. (\ref{eq:7}), and the fractional abundances in the mixtures were adjusted accordingly.

In this study, I employed the Markov Chain Monte Carlo (MCMC) simulation technique \citep{hogg2018data} to estimate the model abundances using Equations (\ref{eq1}) and \ref{eq:7}), following the approach outlined in \cite{emran2021thermophysical}. The MCMC technique used here adopts a Bayesian inference framework to predict the model parameters, e.g., the abundances of each endmember. Using the \textit{emcee} Python package \citep{foreman2013emcee}, I implemented the models and performed 1,000 iterations to estimate the parameters.

\section{Results}
\begin{figure*}
	\centering 
	\includegraphics[width=1.\textwidth, angle=0]{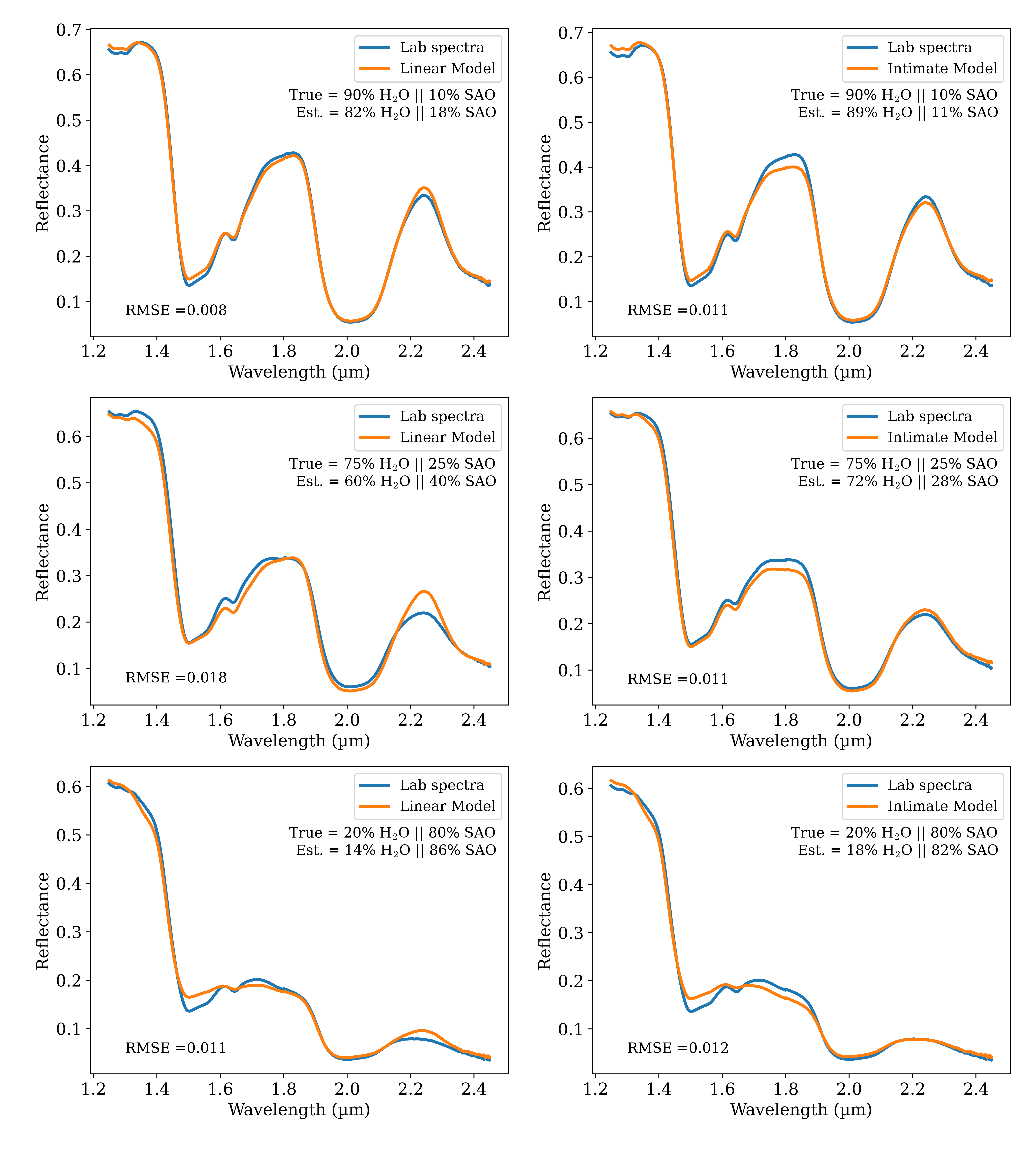}	
	\caption{Comparison of linear mixture modeling (left panel) and intimate mixture modeling using the \cite{hapke1981bidirectional} radiative transfer theory (right panel). In each subplot, the blue spectrum represents the laboratory reflectance spectrum of the H\textsubscript{2}O–SAO mixture \citep{hayes2025insights}, while the orange spectrum shows the modeled spectrum from either the LM or Hapke-based RT modeling. The true (laboratory) and estimated (model-derived) mean abundances of H\textsubscript{2}O ice and sulfuric acid octahydrate (SAO) are indicated in each subplot, along with the root mean square error (RMSE) of the spectral fit.} 
	\label{fig2}%
\end{figure*}

\begin{figure*}
	\centering 
	\includegraphics[width=1.\textwidth, angle=0]{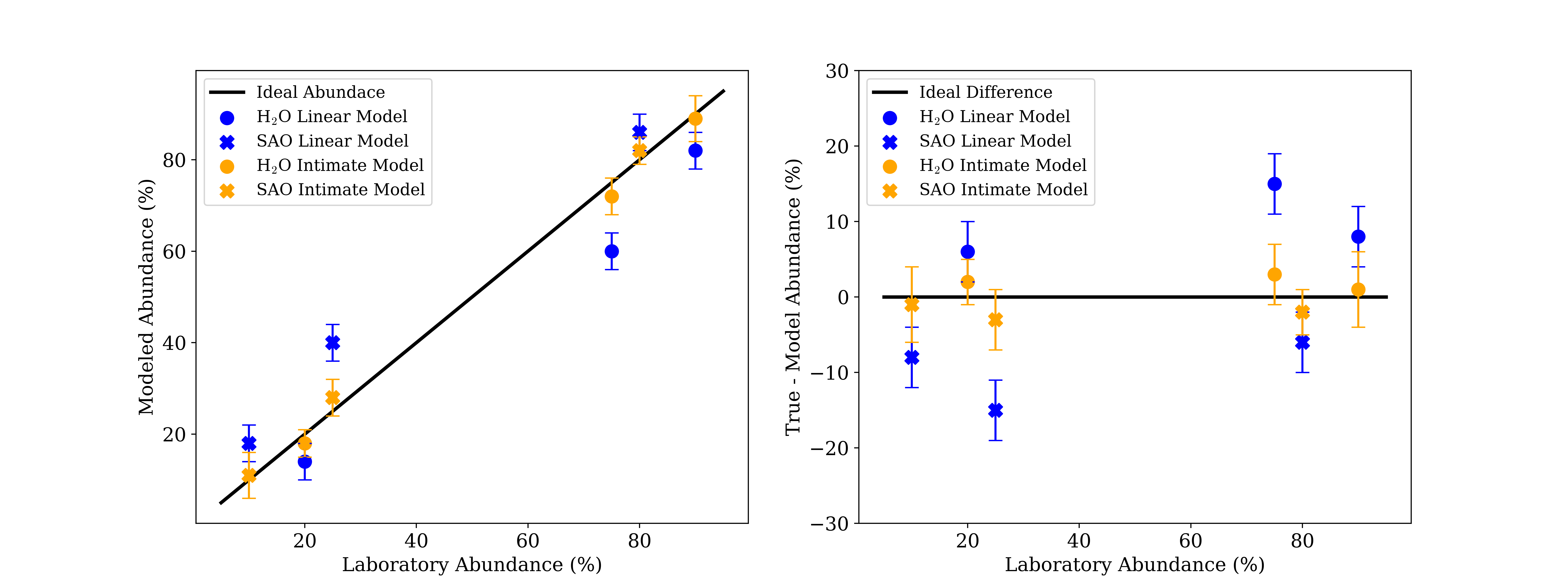}	
	\caption{Comparison of modeled abundances (mean ± 1$\sigma$) of H\textsubscript{2}O ice and sulfuric acid octahydrate (SAO) using the linear mixture modeling and the intimate mixture modeling based on the \cite{hapke1981bidirectional} radiative transfer theory. \textit{Left panel}: Modeled abundances plotted against the expected (ideal) laboratory abundances. Blue and orange markers represent estimates from the LM and Hapke-based RT modeling, respectively. Circle and cross markers represent H\textsubscript{2}O ice and SAO abundances, respectively, by the corresponding mixture modeling. \textit{Right panel}: Difference between the laboratory (true) abundances and the modeled abundances (True-Model abundance) for each component using both modeling approaches (LM and RT).} 
	\label{fig3}%
\end{figure*}

Fig. \ref{fig2} presents the modeled spectra generated using the linear mixture (left panel) and radiative transfer (right panel) modeling approaches, along with the estimated best-fit mean abundances (Est.) for H\textsubscript{2}O ice and H\textsubscript{2}SO\textsubscript{4}$\cdot$8H\textsubscript{2}O across different laboratory mixtures. The estimated abundances vary with mixture ratios (\%wt), as shown in the row-wise subplots (Fig. \ref{fig2}).

For the 90\% H\textsubscript{2}O $\|$ 10\% SAO mixture (top row in Fig. 2), the RT modeling results in better abundance estimates of H\textsubscript{2}O ice and H\textsubscript{2}SO\textsubscript{4}$\cdot$8H\textsubscript{2}O than the LM modeling, even though the LM modeling produces a lower root mean square error (RMSE) value— 0.008 (LM) vs 0.011 (RT). Specifically, the LM and RT estimated mean ± 1$\sigma$ abundances are 82±4\% H\textsubscript{2}O $\|$ 18±4\% SAO and 89±3\% H\textsubscript{2}O $\|$ 11±3\% SAO, respectively (Table \ref{tab1}). Thus, the mean abundance offset from the true composition is ±8\% for LM and ±1\% for RT modeling (Fig. \ref{fig3}).

In the case of the 75\% H\textsubscript{2}O $\|$ 25\% SAO mixture (middle row in Fig. \ref{fig2}), the abundance estimates by the RT modeling again outperform the LM modeling results. Moreover, the RT modeling also shows a lower RMSE value than the LM modeling— 0.011 vs 0.018, respectively. The estimated abundances are 60±4\% H\textsubscript{2}O $\|$ 40±4\% SAO for LM and 72±4\% H\textsubscript{2}O $\|$ 28±4\% SAO for RT modeling (Table \ref{tab1}), resulting in mean abundance offsets of ±15\% (LM) and ±3\% (RT) from the true values (Fig. \ref{fig3}).

Lastly, for the 20\% H\textsubscript{2}O $\|$ 80\% SAO mixture (bottom row in Fig. \ref{fig2}), both models perform comparably in terms of RMSE value (0.012 by RT and 0.011 by LM), though the RT modeling still results in closer abundance estimates to the true abundance. The LM and RT estimates are 14±4\%H\textsubscript{2}O $\|$ 86±4\% SAO and 18±5\% H\textsubscript{2}O $\|$ 82±5\% SAO, respectively (Table \ref{tab1}), corresponding to mean abundance offsets of ±6\% and ±2\%, respectively (Fig. \ref{fig3}).

\begin{table*}[ht]
\centering
\renewcommand{\arraystretch}{1.2}
\caption{Modeled abundances of H$_2$O ice and sulfuric acid octahydrate (SAO) using linear mixture modeling and intimate mixture modeling based on the \cite{hapke1981bidirectional} radiative transfer theory. Abundances are reported as fractional percentages (\%wt) as mean ± 1$\sigma$.}
\label{tab1}
\small
\begin{tabular}{lcccc}
\hline
\makecell{\textbf{Mixture abundance} \\ \textbf{(laboratory mixture)}} &
\multicolumn{2}{c}{\makecell{\textbf{Linear mixture} \\ \textbf{(LM) abundance}}} &
\multicolumn{2}{c}{\makecell{\textbf{Intimate mixture} \\ \textbf{(RT) abundance}}} \\
\cline{2-5}
 & H$_2$O (\%) & SAO (\%) & H$_2$O (\%) & SAO (\%) \\
\hline
90\% H$_2$O + 10\% SAO & 82 ± 4 & 18 ± 4 & 89 ± 3 & 11 ± 3 \\
75\% H$_2$O + 25\% SAO & 60 ± 4 & 40 ± 4 & 72 ± 4 & 28 ± 4 \\
20\% H$_2$O + 80\% SAO & 14 ± 4 & 86 ± 4 & 18 ± 5 & 82 ± 5 \\
\hline
\end{tabular}%

\end{table*}

\newpage
\section{Discussion and implications for Europa}
In this study, I compared two widely-used spectral modeling approaches—
linear mixture modeling and radiative transfer–based intimate mixture modeling —against laboratory spectra of H\textsubscript{2}O ice and H\textsubscript{2}SO\textsubscript{4}$\cdot$8H\textsubscript{2}O mixtures at varying ratios (\%wt) relevant to Europa \citep{hayes2025insights}. The results suggest several key insights regarding the future implementation of spectral modeling (LM and RT) for constraining surface compositional abundances, and by extension, for interpreting the geological processes and surface compositional evolution on Jupiter's moon. 

This study demonstrates that the spectral fit quality—quantified by metrics such as root mean square error (RMSE) or chi-square ($\chi$\textsuperscript{2}) values—does not necessarily reflect the accuracy of the estimated compositional abundances. For example, in the 90\% H\textsubscript{2}O $\|$ 10\% SAO mixture case, the LM model resulted in a lower RMSE compared to the RT model (also, the LM visually shows a closer spectral fit to the laboratory spectra), yet the RT approach produced significantly more accurate abundance estimates. Conversely, in another case (75\% H\textsubscript{2}O $\|$ 25\% SAO mixture), a better spectral fit—as indicated by a lower RMSE—did correspond to abundance estimates that closely matched the true compositional abundance. This suggests that while a lower RMSE or $\chi$\textsuperscript{2} value may sometimes reflect improved abundance accuracy, it is not a consistent indicator or does not guarantee the model's performance in compositional retrieval (abundance estimates of surface materials) on Europa.

\begin{figure*}
	\centering 
	\includegraphics[width=1.\textwidth, angle=0]{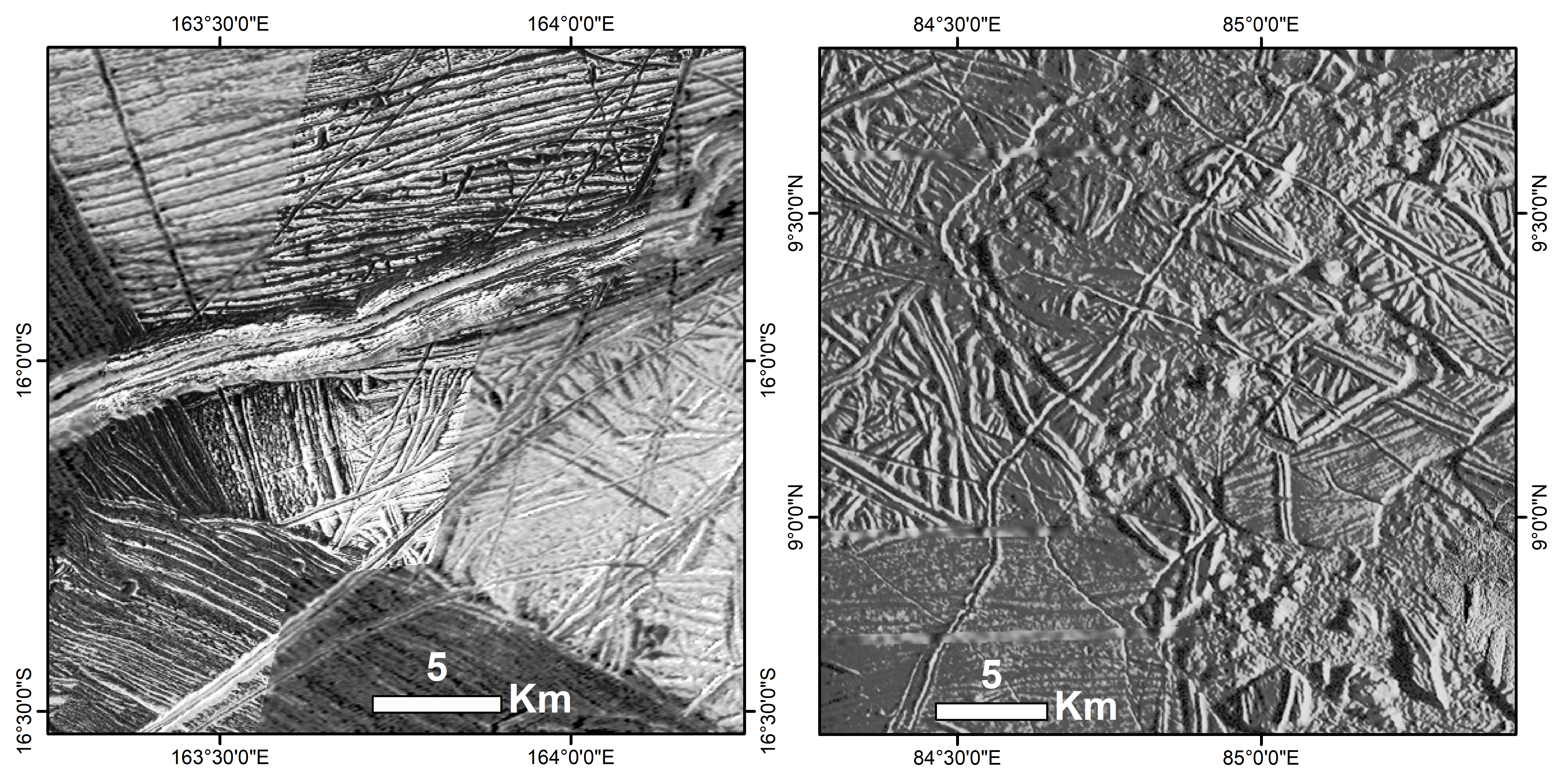}	
	\caption{\textit{Left panel:} Patches of dark and bright materials on Europa observed in a high-resolution image from the Galileo Solid State Imaging (SSI) system \citep{belton1992galileo}. The image is centered at approximately 16°S, 163°45'E, with a spatial resolution of $\sim$50 m/pixel. \textit{Right panel:} A high-resolution Galileo/SSI image of Europa’s chaos terrains on the trailing hemisphere (near center)—an area exposed to radiolytic processes \citep[e.g.,][]{carlson1999sulfuric, carlson2002sulfuric}— showing no discernible dark and bright patches. The image is centered at approximately 9°15'N, 84°50'E, with a spatial resolution of $\sim$50 m/pixel. Both SSI images are adopted from \cite{malaska2024updated}.} 
	\label{fig4}%
\end{figure*}

Across all tested mixture ratios, the RT modeling consistently produces mean abundance estimates that are within ±5\% of the true laboratory values for H\textsubscript{2}O ice and SAO mixtures with $\sim$100 $\mu$m grains. In contrast, the LM modeling tends to result in larger deviations, typically in the range of ±5–15\%. This observation may suggest that Europa’s surface is more accurately represented by an intimate mixture (RT) modeling, regardless of geologic units \citep[e.g.,][]{greeley2000geologic, prockter2009morphology, daubar2024planned, leonard2024global} across the leading and trailing hemispheres, given that the grain sizes are comparable ($\sim$100 $\mu$m). However, the deviation in the LM estimates is relatively smaller when sulfuric acid abundance dominates in the mixture. For instance, the LM estimated abundances deviate by only ±6\% from the true values in a high-SAO scenario (80\% in mixture). A similar deviation in abundance (±8\%) is also observed where sulfuric acid is present in trace amounts, such as in $\sim$10\% of the mixture. Thus, in regions where surface materials show a patchy distribution and the abundance of sulfuric acid is either very high (\(>80\%\)) or very low ($\sim$10\%), the LM model produces viable abundance estimates with $\sim$100 $\mu$m grains. For instance, a surface composed of patches of dark and bright materials on Europa (Fig. \ref{fig4})— arranged in a checkerboard-like distribution— may favor the application of the LM model \citep[e.g.,][]{shirley2016europa}, as this approach is based on areal mixing of materials. In contrast, RT modeling is better suited for regions where surface materials are thoroughly mixed at the particle level— similar to a lunar regolith–type intimately mixed surface \citep[e.g.,][]{shkuratov1999model}— rather than being segregated into distinct patches, which could result from thermal segregation of surface materials \citep{spencer1987thermal, sorli2023thermal}. RT modeling may particularly be relevant in areas influenced by radiolytic processes on Europa \citep[Fig. \ref{fig4};][]{shirley2016europa}, where radiation likely interacted uniformly with the molecules in the surface regolith of the moon \citep{carlson1999sulfuric,carlson2002sulfuric}. Nonetheless, regardless of the magnitude of deviation, both LM and RT modeling consistently overestimates the abundance of  H\textsubscript{2}SO\textsubscript{4}$\cdot$8H\textsubscript{2}O and underestimating H\textsubscript{2}O ice (Table \ref{tab1}). These discrepancies are consistently observed across all laboratory mixture cases analyzed in this study (see Fig. \ref{fig3}).

Photons at NIR wavelengths penetrate a finite depth into a particulate surface \citep[e.g.,][]{born1999principles}. This penetration depth is a critical factor, as it governs multiple scattering effects in intimate mixtures or granular media \citep{hapke1981bidirectional, hapke2012theory}. The areal mixing approach assumes that the surface is composed of patches with uniform composition and does not account for variations in depth within grains or regolith. Under this assumption, each photon interacts with only a single endmember, thereby neglecting multiple scattering between different compositional units or grains. In contrast, intimate mixtures exhibit highly nonlinear spectral behavior at NIR wavelengths because light is multiply scattered among mineral species or grains \citep{hapke1981bidirectional, hapke2012theory, mustard1987quantitative, mustard1989photometric}. As a result, even trace amounts of material can significantly influence the observed spectra of an intimate mixture \citep[e.g.,][]{hapke2012theory, hayes2025insights}. These limitations of the LM modeling approach— which ignores photon penetration depth, multiple scattering, and compositional variation with depth— likely contribute to the higher discrepancies in abundance estimates of mixture components, particularly when grain sizes are much smaller than the photon penetration depth. By comparison, RT modeling accounts for photon penetration depth by incorporating multiple scattering effects when estimating the complex spectral properties of mixtures. While multiple scattering strongly influences reflectance spectra used in LM modeling, its effects are reduced when reflectance is converted to single scattering albedo \citep[e.g.,][]{hapke1981bidirectional, hapke2012theory}. In this form, the single scattering albedo spectra of a mixture is approximately linearly proportional to the relative fractional abundances of the constituent materials \citep[e.g.,][]{hapke1981bidirectional, hapke2012theory, mustard2019theory}. Thus, by properly accounting for the complexity of light scattering in granular or particulate media, RT modeling provides more accurate abundance estimates—within ±5\% in this study— compared to LM modeling.

The performance of the LM and RT modeling was compared here using laboratory spectra of H\textsubscript{2}O ice and sulfuric acid with grain sizes ranging from 90–106 $\mu$m \citep{hayes2025insights}. Due to the limited availability of laboratory spectra for other grain sizes for H\textsubscript{2}O and H\textsubscript{2}SO\textsubscript{4}$\cdot$8H\textsubscript{2}O mixtures, the influence of grain size on the modeling performance for these mixtures could not be directly assessed. However, the effect of grain size on reflectance spectra—and thus on spectral modeling—cannot be overlooked, as it is well-established that grain size significantly affects spectral features \citep[e.g.,][]{stephan2021vis, mastrapa2008optical, emran&chevrier2023}. Relatively small grain sizes—ranging from 20–30 $\mu$m on the leading hemisphere to $\sim$100 $\mu$m near the poles of the trailing hemisphere—have been suggested to characterize Europa’s surface \citep[e.g.,][]{kieffer1974frost, hansen2004amorphous, carlson2005distribution, carlson2009europa, moore2009surface, becker2024exploring}. However, spectral analyses suggest the presence of coarser grains—up to $\sim$1 mm in size—particularly in regions such as large lineae, microchaos, and band geologic units \citep{ligier2016vlt, emran2025nh}. Thus, the findings presented here are most applicable to regions on Europa where grain sizes are relatively small, around $\sim$100 $\mu$m. In contrast, terrains associated with effusive cryovolcanism (viz. surface flow) or surface–subsurface exchange, such as microchaos, are suggested to host coarser grains, potentially up to $\sim$1 mm in size \citep{ligier2016vlt, emran2025nh}. Consequently, I acknowledge that the results of this study have not been validated for those regions with coarser surface materials. A separate investigation using laboratory spectra with a broader range of grain sizes—including mixtures containing both micron- and millimeter-scale diameter grains—is warranted to assess the impact of grain size on spectral modeling accuracy. Notably, a detailed study comparing the accuracy of LM and RT modeling for H\textsubscript{2}O ice mixtures with coarse ($\sim$1 mm) and fine ($\sim$100 $\mu$m) grains relevant to Europa is presented  in \cite{emran2025paper2}.

Moreover, this study investigated the accuracy of LM and RT modeling using the reflectance spectra of the binary mixtures of H\textsubscript{2}O ice and SAO— the two major components identified on Europa \citep[e.g.,][]{Moroz1966, Pilcher1972, carlson1999sulfuric, carlson2005distribution, carlson2009europa, becker2024exploring}— at different mixing ratios \citep{hayes2025insights}. However, Europa’s surface is also hypothesized to host other species, such as CO\textsubscript{2} \citep{villanueva2023endogenous}, H\textsubscript{2}O\textsubscript{2} \citep{carlson1999b}, SO\textsubscript{2} \citep{lane1981sulphur}, NH\textsubscript{3}-bearing compounds \citep{emran2025nh}, and various hydrated sulfate and chlorine (Cl-bearing) salts (e.g., \citealt{davis2024pwyll}; \citealt{becker2024exploring}, and references therein). To comprehensively assess the performance of LM and RT modeling, it is important to validate these approaches against mixtures that include a broader suite of materials. While laboratory spectra of these species would be invaluable for such testing, well-constrained reflectance spectra of intimate mixtures are limited and currently available for binary mixtures of  H\textsubscript{2}O ice and H\textsubscript{2}SO\textsubscript{4}$\cdot$8H\textsubscript{2}O. Thus, I emphasize the need for a robust set of optical constants and laboratory mixture spectra for these additional constituents, as they are essential for a complete understanding of Europa’s surface composition as constrained through spectral modeling using either areal mixing or radiative transfer approaches.

The use of zero backscattering and isotropic scattering in this study was based on the controlled viewing geometry (phase angle of 30°) of the laboratory spectra and the known grain sizes of the materials. For remotely sensed data, such as from telescopes or spacecraft, however, the scenario differs— viewing geometry varies from observation to observation, and the grain sizes of the regolith are often poorly constrained or unknown. In such cases, photometric analyses of planetary surfaces are performed using the Hapke radiative transfer model \citep{hapke1981bidirectional} to estimate surface backscattering and phase parameters, as done for Europa \citep[e.g.,][]{buratti1985photometry, buratti1995photometry, domingue1992disk, belgacem2018estimation, belgacem2020regional}. Accurate backscattering and phase parameters are therefore essential for constraining reliable abundance estimates of surface compositions  using RT modeling from ground- and space-based observations. 

Nonetheless, both LM and RT models produce acceptable results— within ±10\% uncertainty— under specific compositional conditions (H\textsubscript{2}O and H\textsubscript{2}SO\textsubscript{4}$\cdot$8H\textsubscript{2}O mixtures with grain sizes around $\sim$100 $\mu$m) relevant to regions on Europa. This includes the regions where H\textsubscript{2}SO\textsubscript{4}$\cdot$8H\textsubscript{2}O is the dominant component (\(>80\%\) abundance), such as the trailing hemisphere \citep[e.g.,][]{carlson2005distribution}, as well as regions where H\textsubscript{2}SO\textsubscript{4}$\cdot$8H\textsubscript{2}O occurs in trace amounts ($\sim$10\% abundance), as observed in some parts of the leading hemisphere \citep{dalton2013exogenic, ligier2016vlt}. Overall, the results support the use of radiative transfer-based intimate mixture spectral modeling \citep[e.g.,][]{hapke1981bidirectional, shkuratov1999model} as the preferred method for constraining the surface composition of Europa. However, linear mixture modeling remains a viable approach in compositional regimes where the abundance of sulfuric acid is either very high or very low, allowing for flexibility depending on the available data and modeling constraints. These findings provide a valuable guideline for future compositional analyses using imaging spectrometers, such as ESA-JUICE’s Moons and Jupiter Imaging Spectrometer \citep[MAJIS;][]{poulet2024moons} and NASA-Europa Clipper’s Mapping Imaging Spectrometer for Europa \citep[MISE;][]{blaney2024mapping}, for an improved understanding of the composition of Europa’s surface.

\newpage
\section*{Notations}
\textit{r} = Reflectance or radiance factor

\textit{r\textsubscript{m} } = Reflectance of mixture

\textit{r\textsubscript{o} } = Bihemispherical reflectance for isotropic scatterers

\textit{f} = Fractional abundance/ factor

\textit{w} = Single scattering albedo (SSA)

\textit{w\textsubscript{m}} = SSA of the mixture spectrum

\textit{i } = Incidence angle

\textit{e} = Emission angle

\textit{g} = Phase angle

\textit{B(g)} = Backscattering function

\textit{P(g)} = Single-particle phase function

\textit{H(x)} = Chandrasekhar’s H functions

$\mu$\textsubscript{i} = Cosine of the incidence angle

$\mu$\textsubscript{e} = Cosine of the emission angle

 $\gamma$ = Albedo factor
 
 \textit{$\delta$} = Photon penetration depth

\textit{k} = Imaginary part of optical constants

\textit{d}  = Diameter or Grain Size

\section*{Data Availability}
The laboratory spectral data used in this study were collected from \cite{hayes2025insights} and can be accessed at \href{https://doi.org/10.5281/zenodo.13852483}{https://doi.org/10.5281/zenodo.13852483}. Galileo Solid State Imaging (SSI) data used in this study can be found in the National Aeronautics and Space Administration's Planetary Data System: Imaging Node Server at \href{https://pdsimage2.wr.usgs.gov/}{https://pdsimage2.wr.usgs.gov/} \citep{malaska2024updated}.

\section*{Declaration of generative AI}
During the preparation of this work the author(s) used OpenAI in order to improve the readability and language of the manuscript. After using this tool/service, the author(s) reviewed and edited the content as needed and take(s) full responsibility for the content of the publication.

\section*{Acknowledgments}
This research was carried out at the Jet Propulsion Laboratory (JPL), California Institute of Technology, under a contract with the National Aeronautics and Space Administration (80NM0018D0004). I acknowledge JPL’s High-Performance Computing supercomputer facility, which was funded by JPL’s Information and Technology Solutions Directorate. I also thank Kathryn M. Stack for valuable discussions on spectral modeling. Copyright © 2025. California Institute of Technology. Government sponsorship acknowledged.

\bibliographystyle{elsarticle-harv} 
\bibliography{ref}
\end{document}